\documentclass[runningheads,10pt]{llncs}
\usepackage{amsmath}
\usepackage{amssymb}
\usepackage[T1]{fontenc}
\usepackage{graphicx}
\usepackage{algorithm}
\usepackage{framed}
\usepackage{algpseudocode}
\usepackage{enumitem}
\usepackage{url}

\newcommand\adversary{\mathcal{A}}
\newcommand\challenger{\mathcal{C}}
\newcommand\simulator{\mathcal{S}}
\newcommand\extractor{\mathcal{E}}

\newcommand{\GVf}{\textit{GVf}}
\newcommand{\UVf}{\textit{UVf}}
\newcommand{\WReg}{\textit{WReg}}
\newcommand{\RReg}{\textit{RReg}}
\newcommand{\USK}{\textit{USK}}
\newcommand{\CrptU}{\textit{CrptU}}
\newcommand{\AddU}{\textit{AddU}}
\newcommand{\SndToI}{\textit{SndToI}}
\newcommand{\SndToU}{\textit{SndToU}}
\newcommand{\Sig}{\textit{Sig}}
\newcommand{\Trace}{\textit{Trace}}

\begin{document}
\title{Nicknames for Group Signatures}
%
%
\author{Guillaume Quispe\inst{1} \and
Pierre Jouvelot\inst{1,2}
\and
Gérard Memmi\inst{1}
}
\authorrunning{G. Quispe et al.}
%
\institute{LTCI, Télécom Paris, IP Paris, Palaiseau, France  \and
Mines Paris, PSL University, Paris, France  
\email{firstname.lastname@telecom-paris.fr}\\
}
\maketitle              
\begin{abstract}
     Nicknames for Group Signatures (NGS) is a new signature scheme that extends Group Signatures (GS) with Signatures with Flexible Public Keys (SFPK). Via GS, each member of a group can sign messages on behalf of the group without revealing his identity, except to a designated auditor.
     Via SFPK, anyone can create new identities for a particular user, enabling anonymous transfers with only the intended recipient able to trace these new identities.  
     To prevent the potential abuses that this anonymity brings, NGS integrates flexible public keys into the GS framework to support auditable transfers.
     
        In addition to introducing NGS, we describe its security model and provide a mathematical construction  proved secure in the Random Oracle Model.  
        As a practical NGS use case, we build NickHat, a blockchain-based token-exchange prototype system on top of Ethereum. 
    
\end{abstract}

\section{Introduction}
    \label{sec:introduction}

Privacy is a fundamental property that safeguards users’ information and communications. 
It is also enforced by a large number of legal frameworks worldwide such as the General Data Protection Regulation \cite{gdpr2016general} in Europe.
    A variety of cryptographic schemes, or ``primitives'', aim at providing different solutions to enable anonymity and  privacy. 
    For instance, blind signatures~\cite{Chaum1982BlindSF} conceal the message being signed, while ring signatures~\cite{rivest2001leak} hide the signer's identity within the ring he creates.
    Signatures with Flexible Public Keys (SFPK)~\cite{backes_signatures_2018} allow anyone to designate a user anonymously by creating a new identity for him.
    
    But full anonymity may not always be desirable as it can protect abuses and malicious activities. In many cases, some control is required. 
    To this end, primitives such as traceable ring signatures \cite{fujisaki2007traceable} sanction users if they ever misbehave, e.g., when they vote more than once, by revealing their identity; in other words, cheaters will be exposed. 
    Group signatures (GS) \cite{chaum1991group} offer a stronger control by introducing an opener as the only entity able to unveil the identity of a member by ``opening'' his signature; this signature only reveals his belonging to the group.

    Following GS, Kiayias et al~\cite{kiayias2007group} add control on encryption and introduce Group Encryption (GE), a primitive that hides the identity of a ciphertext recipient while still guaranteeing his belonging to the group. 
    Yet, an opening authority can reveal the recipient's identity, i.e., the member able of decrypting the message. 
    Instead of encryption, our paper focuses on ownership by introducing control over flexible public keys, i.e., SFPK, thus enabling auditable transfers, a feature particularly relevant in the financial sector, among others. 

    \paragraph{Nicknames for Group Signatures}
    We introduce a new scheme called ``Nicknames for Group Signatures'' (NGS) that lets anyone 
    generate a new identity, called ``nickname'', for a group member, while still ensuring verifiable group membership of this new identity.  
    The targeted member can then retrieve and prove ownership of this new nickname. 
    This identity cannot be linked to the targeted group member, except by the latter and an opening authority.
       More generally, one can see a nickname as a kind of pseudonym for a member, but, and this is key, assigned by others rather than chosen by himself.
NGS thus merges concepts from GS and SFPK and can be regarded as an extension of group signatures. 

    We believe that NGS is the first signature scheme in which users or entities must be registered as part of a managed group while their interactions or connections are both concealed from the public and auditable. This concealment is strong enough to prevent even correlations between interactions to be inferred.
    
    Moreover, NGS and its security model have been designed so as to yield a generic signature solution that can be used in other similar situations without having to introduce new, application-specific security properties.  
    Activities that could benefit from NGS include payment systems, 
    supply-chain management~\cite{chiacchio2022non}, crowdfunding~\cite{wang2019crowdchain}, or car sharing systems~\cite{valavstin2019blockchain}.
 In fact,     
    NGS has already been put to practical use
    within two blockchain-based applications built on top of Ethereum~\cite{Eth14}. NickPay~\cite{quispe2025nickpay} 
    is a payment system in similar vein as in \cite{AndroulakiCCDET20,BLKZ25,TBAAGPY22}, and NickHat is a new token-exchange system described in  Appendix~\ref{sec:application}. 
\paragraph{Construction}
    We also introduce a NGS construction, i.e., a precise mathematical definition of the NGS concepts, focused on efficiency, since we are  targeting practical and scalable financial applications on Ethereum (see Appendix \ref{sec:application}).
    Our construction of NGS builds upon Signatures with Efficient Protocols (SEP)~\cite{camenisch2002signature}, i.e., secure signature schemes that allow signature issuance on committed messages and proof of knowledge of a signature. This approach has led to efficient GS schemes~\cite{kim2021practical}, based on randomizable signatures~\cite{getshorty}\cite{shortrandomizable}, from which we derive our NGS construction. 
    We prove this construction secure in the Random Oracle Model (ROM) for the NGS security model we propose.

    \paragraph{Contributions}
    In this paper, we thus introduce the following contributions:
    \begin{itemize}
        \item the new Nicknames for Group Signatures scheme (NGS), its formal definition and its correctness property;
        \item the  GS security properties 
         that NGS extends (non-framea\-bi\-lity, traceability, optimal opening soundness, ano\-ny\-mity) and a new one, opening coherence;
        \item a mathematical construction of NGS, based on \cite{kim2021practical}, with a thorough  analysis of its security properties;
        \item an NGS application, NickHat, a new token-exchange application on Ethereum.
        \end{itemize}

        For space reasons, we do not describe here the new Rust computer implementation of the NGS construction used, in particular, in NickHat.
        A short version of this paper is published in \cite{QJM26}.

    \paragraph{Organization}
    After this introduction, we briefly present, for self-containedness, in Section~\ref{sec:background}, background information regarding signatures and proof protocols. Then, Section~\ref{sec:related} covers work related to cryptographic schemes addressing privacy issues.
    In Section~\ref{sec:ngs}, we introduce NGS along with its security model.
    We then describe, in Section~\ref{sec:construction}, an efficient NGS construction based on~\cite{kim2021practical}, together with its associated security proofs (see Appendix~\ref{sec:proofs}).
    Finally, Section~\ref{sec:conclusion} concludes and offers some perspectives for possible future work. Appendix~\ref{sec:application} describes NickHat, an illustrative  NGS-based blockchain application. 

\section{Background}
\label{sec:background}
\subsection{Digital signature}\label{sec:signature}
    \begin{definition}[Digital signature]
        A {\em digital signature scheme} (DS) consists of 3 algorithms.
        \begin{itemize}
            \item $\text{KeyGen}$ is a key generation algorithm that outputs a pair of secret and public keys, when given a security parameter.
            \item $\text{Sig} $ is a signing algorithm taking a secret key and a message $m$ and yielding the signature for $m$.
            \item $\text{Vf}$ is a verification algorithm that takes a signature $\sigma$, a message $m$ and a public key $pk$ and outputs 1, if $(m,\sigma)$ is valid (i.e., the secret key used to build $\sigma$ and $pk$ match under DS), and 0, otherwise.
        \end{itemize}
    \end{definition}

    The standard security property for a digital signature scheme is ``existential unforgeability
    under chosen message attacks'' (EUF-CMA) \cite{goldwasser1988digital}.

    \subsection{Group signature}

    In group signatures, only the members of a group (also called ``{signers}''), managed by an ``issuer'' and an ``opener'', can sign messages on behalf of the group, providing thus anonymity.
    Moreover, a tracing authority, the {opener}, can revoke the anonymity of the signer.
    \begin{definition}[Group signature]
        A {\em group signature scheme} (GS) is a tuple of probabilistic polynomial-time (PPT) algorithms (Setup, IKg, OKg, UKg, GJoin, GSig, GVf, GOpen, GJudge) defined as follows.
        \begin{itemize}
            \item Given a security parameter, Setup  outputs a public parameter.
            \item $IKg,OKg$ and $UKg$ allow, given a public parameter, the issuer, opener, and user to invoke their respective key generation algorithm to output their  secret/public key pair.
            \item $GJoin~ (Join \leftrightarrow Iss)$ allows a user to join the group by an interactive protocol between the user and the issuer.
            The $Join$ algorithm is run by the user, the $Iss$ algorithm by the issuer.
            If successful, the user receives his group signing key and the issuer adds the user's information on a registration list.
            \item $GSig$ allows a user, when owning a group signing key, to output a signature on a message on behalf of the group.
            \item $\textit{GVf}$,  publicly available, takes an issuer public key, a message, and a group signature and outputs 1, if the signature  is valid with respect to the issuer public key, and $0$, otherwise.
            \item $GOpen$ is called by the opener with his opener secret key to identify the signer of a group signature, outputting the index of the user with a proof stating that he did produce the signature or $\perp$, if no user has been found.
            \item $GJudge$ can be used to verify an opener's proof.
            It takes a message, a group signature on the message, the issuer public key, a user index with her public key and the opener's proof to be verified.
            It outputs $1$, if the proof is valid, $0$, otherwise.
        \end{itemize}
    \end{definition}
    As for DS, GS and its variant with groups of varying sizes are ``correct'' when a signature from an honest member can be verified  with probability 1. 
    
    GS supports a security model that we very briefly and informally describe here, following Bellare et al~\cite{bellare2005foundations}. 
    ``Anonymity'' ensures that no adversary can identify the signer of a group signature.
    ``Non-frameability'' ensures that an honest user cannot be falsely accused of having signed a message.
    ``Traceability'' ensures that no user can produce a valid group signature that is not traceable by the opener.
    ``Opening soundness'' ensures that no adversary can produce a signature that can be opened to two distinct users.
    Note that other security properties such as unforgeability had been previously introduced, but Bellare et al.~\cite{bellare2005foundations} showed that their four properties, sketched above, encompass these.

    \subsection{Signature with flexible public key}

        In this scheme, the public key of a user belongs to an equivalence class induced by a relation $\mathcal{R}$. Anyone can change it while maintaining it in its equivalence class.
    Only through the use of a ``trapdoor'' can one check that a given public key is member of an equivalence class.
    \begin{definition}[Signature with flexible public key]
        A {\em signature scheme with flexible public key}
        (SFPK) is a tuple of PPT algorithms (KeyGen,TKGen, Sign, ChkRep, ChgPK, Recover, Verify) is defined as follows.
        \begin{itemize}
            \item $KeyGen$ is the key-generation algorithm that takes as input a security parameter and random coins and outputs a pair of secret and public keys.
            \item $TKGen$ is the trapdoor generation algorithm that takes as input a security parameter and random coins and outputs a pair  of secret and public keys with its corresponding trapdoor.
            \item $Sign$ is the signing algorithm that takes as input a secret key  and a message to sign and outputs a signature  valid for the message.
            \item $ChkRep$ takes as input a trapdoor for some equivalence class and a public key and outputs 1, if the public key is in the equivalence class, and 0, otherwise.
            \item $ChgPK$ takes as input a representative public key of an equivalence class and a random coin and returns a different representative public key from the same equivalence class.
            \item $Recover$ takes as input a secret key, a trapdoor, and a representative public key and returns an updated secret key.
            \item $\textit{Verify}$ takes as input a public verification key, a message and a signature and outputs 1, if the signature is valid for the message, and 0, otherwise.
        \end{itemize}
    \end{definition}

    \subsection{Proof protocols}\label{subsec:sigma}
    We recall the notion of $\Sigma$ protocols and the notation from \cite{getshorty}.
    Let $\phi: \mathbb{H}_1 \rightarrow \mathbb{H}_2 $ be an homomorphism with $\mathbb{H}_1$ and $\mathbb{H}_2$ being two groups of order $q$ and let $y \in \mathbb{H}_2$. 

    We denote by $PK\{ (x):y=\phi(x)\}$ the $\Sigma$ protocol for a zero-knowledge proof of knowledge of an $x$ such that $y=\phi(x)$.
    $\Sigma$ protocols are three-move protocols between a prover $P$ and a verifier $V$ described as follows.
    \begin{enumerate}
        \item $P \rightsquigarrow V$: $P$ chooses $\textit{rnd} \leftarrow_\$\mathbb{H}_1$ and sends $\textit{Comm} = \phi(\textit{rnd})$ to $V$, where $\leftarrow_\$$ denotes the uniform-sampling operation on a finite set, here on $\mathbb{H}_1$.
        \item $V \rightsquigarrow P$: Once $\textit{Comm}$ is received, $V$ chooses a challenge $\textit{Cha} \leftarrow_\$ \mathbb{H}_1$ and sends it to $P$.
        \item $P \rightsquigarrow V$: $P$ computes $\textit{Rsp}= \textit{rnd}-\textit{Cha} \cdot x$, and sends it to $V$, who checks whether ($\phi(\textit{Rsp})\cdot\phi(x)^{\textit{Cha}}=\textit{Comm})$ or not.
    \end{enumerate}

    We denote by $\pi=SPK\{(x): y=\phi(x)\}(m)$, with $m \in \{0,1\}^*$, the signature variant of a $\Sigma$ protocol obtained by applying the Fiat-Shamir heuristic~\cite{Fiat87} to it; this is called a ``signature proof of knowledge'' on a message $m$.
    The Fiat-Shamir heuristic, below, removes the interaction in Steps 2 and 3 above by calling a ``random oracle'' to be used in security proofs, instantiated by a suitable hash function $H:\{0,1\}^* \rightarrow \mathbb{Z}_q$.

    \begin{enumerate}
        \item $P \rightsquigarrow V$: $P$ chooses $\textit{rnd} \leftarrow_\$ \mathbb{H}_1$, computes $\textit{Cha} = H(\phi \mathbin\Vert y \mathbin\Vert \phi(\textit{rnd}) \mathbin\Vert m)$ and $\textit{Rsp}=\textit{rnd}-\textit{Cha}\cdot x$, and sends $(\textit{Cha}, \textit{Rsp})$ to $V.$
        \item $V$: $V$ accepts $(\textit{Cha}, \textit{Rsp})$ iff ($\textit{Cha}= H(\phi\mathbin\Vert y\mathbin\Vert\mathbin\Vert y^{\textit{Cha}}\cdot \phi(\textit{Rsp})\mathbin\Vert m))$.
    \end{enumerate}
    \noindent where $\mathbin\Vert$ is the string-concatenation function on binary numbers (we assume a proper binary encoding of the $\phi$ function and of elements of $\mathbb{H}_2$).

    The security properties of SPKs, previously defined in~\cite{groth2006simulation}, can be informally described as follows.
         ``Completeness'' states that a signature generated by an honest signer should be verified successfully.
         ``Zero-knowledge'' (ZK) ensures that a zero-knowledge simulator $\simulator$ able to simulate a valid proof, a SPK, without knowing the witness $x$ and indistinguishable from a real one, does exist.
         ``Simulation soundness'' (SS) states that malicious signers with no witness cannot generate  proofs for false statements (even after receiving simulated proofs).
        \``Simulation-sound extractability'' (SE) ensures that there exists a knowledge extractor $\extractor$ able to extract a correct witness from a valid proof generated by a malicious signer.

    \begin{definition}[Simulation-sound extractable SPK]
        A protocol is a {\em si\-mu\-lation-sound extractable} non-interactive zero-knowledge (NIZK) signature variant of a $\Sigma$ protocol (SPK) if it satisfies completeness, zero-knowledge, and simulation-sound extractability.
    \end{definition}
    \section{Related work}
    \label{sec:related}
    \noindent 
    We focus here on generic cryptographic schemes close to NGS and refer the reader to 
    \cite{quispe2025nickpay} 
    for payment-focused related work with tailored schemes and security models. 

        In~\cite{abe2013double}, the authors propose a public-key anonymous tag system, where users can create tags with their private key, thus for themselves, and an authority can publish a trapdoor (or ``tracing token'') that publicly links all the tags related to one user without revealing who he is. 
    But the same trapdoor can be computed by this user, enabling him to then prove or deny the link between a tag and his public key. 
    As in NGS, the authority and the user share the same trapdoor, but no means to contact users are provided in~\cite{abe2013double}. 
    
    With traceable signatures~\cite{kiayias2004traceable}, the concern was about protecting honest users' anonymity while still enabling the identification of malicious ones.
    A trusted party can ask a group manager to investigate on suspect signatures, and the latter returns a trapdoor of the suspect, allowing the trusted party to open all the signatures of the suspect members while the anonymity of the other members is still protected. 
    They also introduced the self-traceability property, where a member is able to claim that a signature is his. This approach can be seen as group signatures with additional anonymity management, however again with no way to reach a member. 

    With pseudonym systems~\cite{lysyanskaya2000pseudonym} and anonymous credentials, users ask organizations for a credential on attribute(s) so that they can claim possession of them anonymously. 
    To register with an organization, the user and the organization together usually compute a pseudonym for the former. 
    In \cite{camenisch2001efficient}, the first anonymous credential with optional anonymity revocation is introduced.  
    In~\cite{hebant2023traceable}, the authors build an anonymous-credentials system based on two primitives they introduce. 
    With the first one, which concerns us here, named {EphemerId}, users are associated with a tag pair, a secret and its corresponding public tags, which they can randomize and derive the witness for.
    They also extend this primitive with a tracing authority that can identify users given any public randomized tag. 
    However, they keep the general idea underpinning anonymous credentials, thus not offering a transfer system that would be both private and auditable.

    \section{Nicknames for Group Signatures}\label{sec:ngs}

    We describe Nicknames for Group Signatures (NGS), a new digital signature scheme that adds the concept of ``nicknames'' on top of the notion of group signatures. We start by the formal interface and then define the protocol. 
    
    We use equivalence classes in this definition to enable a seamless integration with SFPK.
    Let $p$ denote the modulus of a scalar field, $\mathbb{Z}_p$, appropriate for a given security parameter $\lambda$, and $l$ the dimension of the space $\mathbb{G^*}^l$ of public keys $pk$, represented as tuples, where 
    $\mathbb{G}$ is some cyclic group of large prime order $q$.
    We introduce an equivalence relation $\mathcal{R}$ that partitions $\mathbb{G^*}^l$  into equivalence classes:
    $\mathcal{R}=\{({pk_1},{pk_2}) \in (\mathbb{G^*}^l)^2~ | ~\exists s \in \mathbb{Z}_p^*, {pk_1}={pk_2}^s \},$
    \noindent where exponentiation of tuples is defined element-wise.

    \subsection{Interface}
    \label{sec:interface}

    In the NGS scheme, each participant has a role that provides him with rights to handle particular data or variables and abilities to perform parts of the NGS scheme. 
            A User becomes a {Group Member} by issuing a join request that only can get authorized by the Issuer; then, he will be able to produce nicknames and sign messages. 
            The Opener is the only participant able to ``open'' a nickname to unveil the underlying group member's identity, together with a proof of it; however, anyone can act as a Verifier, to check that a given nickname does exist.
            The Judge role can be adopted to check that a proof supposedly linking a user to a nickname (following its opening)  is valid. 
    One could also define the role of {Group Manager}, which would endorse both roles of issuer and opener. 

To manage these participants, NGS uses specific data types. A Registration information is a structure, usually named $reg$, that
            contains the necessary elements describing a user, saved in the registration table $\mathbf{reg}$ defined below; the exact contents of $reg$ is construction-dependent, but can include, for example, the encryption of the users' trapdoors $\tau$.
 A Join request is a structure, named $reqU$, produced by a user requesting to join the group; $reqU$ contains the necessary construction-dependent elements for a joining request to be handled by the issuer. It can be seen as synchronization data between the user and the issuer during the group-joining process.


    Finally, 
        NGS is built upon a few global variables, which are subject to strict access rights.
        $DS$ is a digital signature scheme vetted by some certification authority, CA. 
            The registration table $\mathbf{reg}$ is  controlled by the issuer, who has read and write access to it; the opener is given a read access to it also.
            The master public key table $\mathbf{mpk}$ is publicly available, and used to define nicknames.
            Finally, the publicly available table $\mathbf{upk}$ stores the users' public keys.

    \begin{definition}[NGS Scheme]
        A {\em nicknames for group signature scheme} (NGS) is a tuple of functions defined in three parts. The first part deals with roles.

        \begin{itemize}
            \item $IKg: 1^\lambda \rightarrow (isk,ipk)$ is the key-generation algorithm that takes
            a security parameter $\lambda$ and outputs an issuer's secret/public key pair ($isk,ipk$).
            \item $OKg: 1^\lambda \rightarrow (osk,opk)$ is the key-generation algorithm
            that takes a security parameter $\lambda$ and outputs an opener's secret/public key pair ($osk,opk$).
            \item $UKg: (i, 1^{\lambda}) \rightarrow (usk,upk)$ is the key-generation algorithm that produces an appropriate $DS$ secret/public key pair for user $i$, given security parameter $\lambda$.
            \item $Join: (usk,ipk,opk) \rightarrow (msk,\tau, reqU)$, the user part of the group-joining algorithm, takes a user's secret key $usk$ and the issuer and opener public keys and outputs a master secret key $msk$ along with a trapdoor $\tau$. $reqU$ will then contain information necessary to the Issuer to run the second part of the group-joining algorithm.
            \item  $Iss: (i,isk,reqU,opk) \rightarrow ()$ or $\perp$, the issuer part of the group-joining algorithm, takes a user $i$, an issuer secret key $isk$, a join request $reqU$ and the opener public key $opk$, updates the registration information $\mathbf{reg}[i]$ and  master public key  $\mathbf{mpk}[i]$, and creates an equivalence class for the user's future nicknames. It returns $\perp$, if registration as a group member fails.
            \end{itemize}
            The second part of NGS deals with nicknames.
            \begin{itemize}
            \item $Nick: (mpk,r) \rightarrow nk$ is the nickname generation algorithm that takes a master public key $mpk$, a random coin r, and creates a nickname $nk$ belonging to $[mpk]_\mathcal{R}$.
            \item $Trace: (ipk, \tau,nk) \rightarrow b$ takes as input the issuer public key $ipk$, the trapdoor $\tau$ for some equivalence class $[mpk]_\mathcal{R}$, and a nickname $nk$ and outputs the boolean $b=(nk \in [mpk]_\mathcal{R})$.
 
            \item $\text{GVf}~: (ipk,nk) \rightarrow b$ is the issuer verification algorithm that takes a issuer public key $ipk$ and a nickname $nk$ and outputs a boolean stating whether nk corresponds to a user member of the group or not.
        
            \item $Open: (osk,nk) \rightarrow (i,\Pi) \text{ or} \perp$,  the opening algorithm, takes an opener secret key $osk$ and a nickname $nk$.
            It outputs the index  $i$ of a user with a proof $\Pi$  that user $i$ controls $nk$ or $\perp$, if no user has been found. 
            \item $Judge: (nk,ipk,i,\Pi) \rightarrow b$ is the judging algorithm that takes a nickname $nk$, an issuer public key $ipk$, 
            a user index $i$ and an opener's proof $\Pi$ to be verified and, using the user public key from the CA $\mathbf{upk}[i]$ for user $i$,  outputs a boolean stating whether the judge accepts the proof or not.
        \end{itemize}
        The last part relates to signatures.
\begin{itemize}
    \item $Sign: (nk,msk,m) \rightarrow \sigma$ takes a nickname $nk$, a master secret key $msk$ and a message $m$ to sign and outputs the signature $\sigma$.
                   \item $\text{UVf} : (nk,m,\sigma) \rightarrow b$ is the user verification algorithm, taking as input a nickname $nk$, a message $m$ and a signature $\sigma$ of $m$ and outputting a boolean stating whether $\sigma$ is valid for $m$ and $nk$ or not. 
\end{itemize}           
                   
    \end{definition}

    Let $m$ be any message that any user $i$, after joining a group, signs with his secret key $msk$ while being nicknamed $nk$. An NGS scheme construction  is said to be ``correct'' when the group opener opens $nk$ to user $i$. Also, the verification functions must validate $nk$ and $(m, \sigma)$, while the proof $\Pi$ generated by $Open$ must be accepted by the $Judge$ algorithm and $nk$ traceable to user $i$, given his  trapdoor $\tau$. This set of conditions is formally defined in Algorithm~\ref{alg:corr}.

        \begin{algorithm}            
        \paragraph{$Exp_{NGS}^{Corr}(\lambda, i, m)$}
            \begin{algorithmic}
                \State $(isk,ipk) = IKg(1^\lambda); (osk, opk) = OKg(1^\lambda)$
                \State $(usk,upk) = UKg(i, 1^\lambda)$
                \State $(msk,\tau, reqU) = Join(usk,opk)  $
                \State $Iss(i,isk, reqU, opk)$
                \State $r \leftarrow_\$ \mathbb{Z}_p; nk = Nick(\mathbf{mpk}[i],r)$
                \State $\sigma = Sign(nk,msk,m)$
                \State $(i',\Pi) = Open(osk,nk)$
                \State {return} $\left( \begin{array}{l}
                (i = i' )
                                            ~\land                           \\
                                            \GVf(ipk, nk) ~\land~
                                            \UVf(nk,m,\sigma) ~\land \\
                                            Judge(nk, i,ipk, \Pi) ~\land~
                                            Trace(ipk, \tau, nk)
                \end{array}\right)$
            \end{algorithmic}
            \caption{Correctness experiment for NGS}
        \label{alg:corr}
        \end{algorithm}
        
    \begin{definition}[NGS Correctness]
        A construction of the NGS scheme is {\em correct} iff  $\textnormal{Pr}[Exp_{NGS}^{Corr}(\lambda, i, m)]=1$ for any parameter $\lambda$, user $i$ and message $m$.
    \end{definition}

\subsection{Protocol}
    We describe here the main steps of a typical use of the NGS interface.

    Creating a new NGS group is performed by selecting both an Issuer and an Opener.
     The issuer authorizes users to become group members and the opener is the only participant able to ``open'' a nickname to unveil the underlying group member's identity.
     The issuer and opener run $IKg$ and $OKg$, respectively, to obtain their pair of secret/public keys. 
    The issuer will then decide (or not) to grant users access to the group he manages, thus enabling the opener  to track, i.e., open, their nicknames used in messages' signatures.
    
    To join the group, a  user, $i$, must first run $\textit{UKg}$ to obtain his secret/public keys $(usk,upk)$ and store $upk$ in the users' public key table as $\textbf{upk}[i]$, vetted by the Certification Authority, CA. 
    Then, he runs the $Join$ function, providing a join request and his secret group-signing key $msk$. 
         Meanwhile, the issuer runs  $Iss$  to verify the correctness of this join request and compute the user's master public key $mpk$, an element of $\mathbb{G^*}^l$. By doing so, he uniquely associates the equivalence class $[{mpk}]_\mathcal{R}$ to user $i$. Finally, he publishes $mpk$ in the master public key table $\mathbf{mpk}$, thus exposing the new member $i$ to everyone.

    Now, NGS enables anonymous transfers by allowing anyone to derive a new identity, a “nickname”, for the recipient $i$.
   To do so, the $\textit{Nick}$ function transforms the recipient's master public key $\mathbf{mpk}[i]$ into a nickname, i.e., a different representative ${nk}$ of the same class $[{mpk}]_\mathcal{R}$, without accessing the secret key controlling ${mpk}$. 

    Anonymity (see below) ensures that the freshly created nickname $nk$ cannot be linked to member $i$, but a verifier can still guarantee that this new identity belongs to the group with the $\textit{GVf}$ function.
    
    Thanks to his NGS-provided trapdoor $\tau$, member $i$ can then detect if  some $nk$ belongs to him, i.e., is in his class $[{mpk}]_\mathcal{R}$, by calling the $\textit{Trace}$ function and profit from the possession of the nickname with the $\textit{Sign}$ function.
    
    Finally, the opener can identify the member hidden behind a nickname with  $\textit{Open}$, and even prove the validity of his results to a judge. 
    The judge can thus being assured that user $i$ indeed controls $nk$ by verifying the proof with  $\textit{Judge}$.
    \subsection{Security model}\label{sec:security}
    GS has been studied for several decades and mathematically formalized (see  Section~\ref{sec:background} and~\cite{bellare2003foundations}).
    To specify NGS security properties, we follow the approach proposed by Bellare et al~\cite{bellare2005foundations}, with the appropriate modifications mentioned below. In particular, to handle nicknames and the existence of \textit{Trace}, a new security property (``opening coherence'') is here introduced.  
    
Informally, when a new nickname identity is created for a recipient, i.e, that he can trace, ``non-frameability'' ensures that no one else can steal and take possession of it. It also protects him from being falsely accused, with the $Open$ algorithm, of having signed a message, as in GS. 
In case of suspicious activity, ``{traceability}'' ensures that the opener can always identify a NGS-signed nickname as a registered user.
To protect user's privacy, the `` {anonymity}'' property ensures that no one can identify a specific signer from a target NGS-signed nickname.
Finally, ``{optimal opening soundness}'' guarantees that no adversary can produce a group-belonging nickname that can be opened to two distinct users,
while the new ``{opening coherence}'' property ensures that the member identified via an opening is the only one able to trace the associated nickname. 

Inspired by~\cite{bellare2005foundations}, we assume  an NGS-based environment set up for one group, with its issuer and opener attacked by PPT adversaries $\adversary^L$ able to take advantage of any  oracle mentioned in the property-specific list $L$. We present, first, the oracles used in our security model.
    Then we formally define, sequentially, the NGS traceability, non-frameability, optimal opening soundness, opening coherence and anonymity properties. 

    \subsubsection{Oracles}
    \label{sec:oracles}

    Figure~\ref{fig:oracles} defines all the oracles that $\adversary$ can use, in addition to a $\textit{Hash}$ function modeled as a  Random Oracle in the security proofs. 
    \begin{itemize}
        \item AddU$(i)$: $\adversary$ uses this oracle to add and then join an honest user $i$.
        \item CrptU($i,upk$): $\adversary$ uses this oracle to corrupt user $i$ and set its public key to be a $upk$ of his choice.
        \item SndToI($i, r$): $\adversary$  uses it to send the join request $r$ of a malicious user $i$  to an honest issuer executing $\textit{Iss}$.
        $\adversary$ does not need to follow the $Join$ algorithm.
        \item SndToU($i$): $\adversary$  uses this oracle to perform the Join part of the group-joining process for an honest user $i$.
        \item USK$(i)$: $\adversary$ uses this oracle to get the secret keys $\mathbf{msk}[i]$ and $\mathbf{usk}[i]$ of an honest user $i$.
        \item RReg($i$): $\adversary$ can read the entry $i$ in the registration table $\mathbf{reg}$ for user $i$.
        \item WReg($i,\rho$): with this oracle, $\adversary$ can write or modify with $\rho$ the entry for user $i$ in the registration table $\mathbf{reg}$.
        \item Trace($i,nk$): $\adversary$ uses this oracle to check if user $i$ owns the provided nickname.
        \item Sig($i,nk,m$): $\adversary$ uses this oracle to obtain a signature on a message $m$ from user $i$ on nickname $nk$.
        \item Ch$_b(i_0,i_1,m)$: for two users $i_0$ and $i_1$ and a message $m$ chosen by $\adversary$, this oracle outputs a NGS-signed nickname $(m, nk,\sigma$) on $m$ under identity $i_b$ as the challenge.
    \end{itemize}

    Below is the set of initially empty lists maintained by the oracles and 
    possibly used by a challenger $\challenger$ in the proofs of the security properties:
    \begin{itemize}
        \item ${L}_h$, honest users with their trapdoors;
        \item ${L}_c$, corrupted users and their current state, $cont$ or $accept$ ($cont$ indicates that the user is corrupted but not yet joined, and $accept$, that the user is corrupted and also accepted to join the system by the issuer);
        \item ${L}_{ch}$, challenged messages and identities, along with the NGS-signed nickname returned by the challenge oracle;
        \item ${L}_{sk}$, user identities queried by  $\adversary$ to access their secret keys via $\USK$;
        \item ${L}_{\sigma}$, queried identities, messages, nicknames and signatures in response to the signing oracle;
        \item ${L}_{tr}$, nicknames, queried by $\adversary$ for users to trace, along with the result. 
    \end{itemize}
     Note that we assumed that three additional data structures were added to the NGS interface, with no change otherwise to its semantics. The tables $\mathbf{usk}$ (updated when a user  calls $\textit{UKg}$) and $\mathbf{reqU}$ and $\mathbf{msk}$ (updated when a user calls $Join$ and $Iss$) keep the user's secret key, request and  master secret key; they are only available within the oracles and $\challenger$.
    \begin{figure*}
        \begin{minipage}[t]{0.5\textwidth}
            \begin{algorithm}[H]
                \paragraph{AddU$(i)$}
                \begin{algorithmic}
                    \State $(usk,upk) = UKg(i, 1^\lambda) $
                    \State $(msk,\tau,reqU) =  Join(usk,opk)$
                    \State $Iss(i,isk,reqU,opk)$
                    \State $L_h = L_h \cup \{(i, \tau)\}$
                    \State return $upk$
                \end{algorithmic}

                \paragraph{SndToI$(i, r)$}
                \begin{algorithmic}
                    \State If $(i,cont) \notin L_c$, return $\perp$
                    \State $L_c = L_c \setminus \{(i,cont)\} \cup \{(i,accept)\}$
        \State return $Iss(i,isk,r, opk)$
                \end{algorithmic}

                \paragraph{SndToU$(i)$}
                \begin{algorithmic}
                    \State $(usk,upk) = UKg(i, 1^\lambda)$
                    \State $(msk,\tau,reqU) = Join(usk,opk)$
                    \State $L_h = L_h \cup \{(i,\tau)\}$
                    \State return $reqU$
                \end{algorithmic}
                \paragraph{Ch$_b$$(i_0,i_1,m)$}
                \begin{algorithmic}
                    \State If $(i_0, *) \notin L_h \lor (i_1, *) \notin L_h$, 
                    \State ~~return $\perp$
                    \State If $(i_0,*,*) \in L_{tr} \lor (i_1,*,*) \in L_{tr}$
                    \State ~~return $\perp$.
                    \State If $i_0 \in L_{sk} \lor i_1 \in L_{sk}$, return $\perp$
                    \State If $\mathbf{mpk}[i_b]$ undefined, return $\perp$
                    \State $r \leftarrow_\$ {\mathbb{Z}_p}; nk  = Nick(\mathbf{mpk}[i_b],r)$
                    \State $\sigma = Sign(nk,\mathbf{msk}[i_b],m)$
                    \State $L_{ch} = 
                    \{(i_0,m,nk,\sigma),(i_1,m,nk,\sigma)\} \cup L_{ch}$
                    \State return $(nk,\sigma)$
                \end{algorithmic}
            \end{algorithm}
        \end{minipage}
        \hfill
        \begin{minipage}[t]{0.5\textwidth}
            \begin{algorithm}[H]

                \paragraph{USK$(i)$}
                \begin{algorithmic}
                    \State If $(i,m,nk,\sigma) \in L_{ch}$ for some $(m,nk,\sigma)$, 
                    \State ~~
                    return $\perp$
                    \State $L_{sk} = L_{sk} \cup \{i\}$
                    \State return $(\mathbf{msk}[i],\mathbf{usk}[i]) $ 
                \end{algorithmic}

                \vspace{1mm}

                \paragraph{RReg$(i)$}
                \begin{algorithmic}
                    \State return $\mathbf{reg}[i]$
                \end{algorithmic}
                \vspace{1mm}

                \paragraph{WReg$(i,\rho)$}
                \begin{algorithmic}
                    \State $\mathbf{reg}[i] = \rho$
                \end{algorithmic}
                \vspace{1mm}
                \paragraph{CrptU$(i,upk)$}
                \begin{algorithmic}
                    \State $\mathbf{upk}[i] = upk$
                    \State $L_{c} = L_{c} \cup \{(i,cont)\}$
                \end{algorithmic}

                \paragraph{Trace$(i,nk)$}
                \begin{algorithmic}
                    \State If $(i,\tau) \notin L_h$ for some $\tau$, return $\perp$
                    \State $b=Trace(ipk, \tau,nk)$
                    \State $L_{tr} = L_{tr} \cup \{(i,nk,b)\}$
                    \State return $b$
                \end{algorithmic}
                \vspace{1mm}
                \paragraph{Sig$(i,nk,m)$}
                \begin{algorithmic}

                    \State If $(i,nk,true) \notin L_{tr}$,
                    return $\perp$
                    \State $\sigma = Sign(nk,\mathbf{msk}[i],m)$
                    \State $L_\sigma = L_\sigma \cup \{(i,m,nk,\sigma)\}$
                    \State return $\sigma$ 
                \end{algorithmic}
                
            \end{algorithm}
        \end{minipage}
        \caption{Oracles for the security model of NGS (the  issuer and opener keys $isk, ipk$ and $opk$ are supposed defined)}
        \label{fig:oracles}
    \end{figure*}

  \subsubsection{Traceability}
    
    In this experiment, the opener is ``partially corrupted'', meaning that the opening must follow the prescribed program but the adversary $\adversary$ is given the opening key $osk$.
    The issuer is kept honest.
    $\adversary$ can call  $\AddU$  to join an honest user, $\CrptU$ and $\SndToI$, to join a corrupt user, and the $\USK$ oracle, to get all users secret keys (the $Sig$ oracle is therefore not needed).
    Finally, $\adversary$ can read the registration list with $\RReg$.

    The adversary wins if he produces a valid NGS-signed nickname from a user who did not join the group, causing the $Open$ algorithm to fail to identify the user\footnote{Note that, since the operations performed in $\textit{Trace}$ are a subset of those of $Open$, this experiment also ensures that an invalid tracing cannot be forged.}.
    He can also win  if a forgery $\sigma$ produced by some user $i$ leads to a proof $\Pi$ from the $Open$ algorithm that gets rejected by the $Judge$ algorithm.

    We define the advantage in the traceability experiment described in Figure~\ref{fig:trace} for any polynomial time adversary $\adversary = \adversary^{\AddU, \CrptU, \SndToI, \USK,\RReg}$ as
    $Adv_{NGS,\adversary}^{Trace}(\lambda)=\textnormal{Pr}[Exp^{Trace}_{NGS,\adversary}(\lambda)=1].$

    \begin{definition}
        A NGS scheme is \emph{traceable} if $Adv_{NGS,\adversary}^{Trace}(\lambda)$ is negligible for any $\adversary$ and a given $\lambda$.
    \end{definition}
    
    \begin{figure}
     
        \begin{algorithm}[H]
      
      \paragraph{$Exp_{NGS,\adversary}^{Trace}(\lambda)$}
            \begin{algorithmic}
                \State
                $(isk,ipk) = IKg(1^\lambda);
                (osk, opk) = OKg(1^\lambda)$
                \State
                $(m^*,nk^*, \sigma^*) = \adversary^{\AddU,\CrptU,\SndToI,\USK,\RReg}(ipk,osk)$
                \State If following conditions hold, then return 1:
                \begin{itemize}
                    \item $\GVf(ipk,nk^*)
                    \land 
                \textit{UVf}(nk^*, m^*,\sigma^*)$;
                    \item
                    $  r = \perp \lor~  \neg Judge(nk^*,i^*,ipk,\Pi), \textnormal{when}~(i^*,\Pi) = r\textnormal{, with}~r = Open(osk, nk^*)$.
                \end{itemize}
                \State return 0
            \end{algorithmic}
\vspace{-3.7pt}
        \paragraph{$Exp_{NGS,\adversary}^{Nf}(\lambda)$}
            \begin{algorithmic}
                \State $(isk,ipk) = IKg(1^\lambda);
                (osk, opk) = OKg(1^\lambda)$
                \State $(m^*,nk^*,\sigma^*,i^*,\Pi^*) = \adversary^{\SndToU,\USK,\Trace,Sig}(isk,osk)$
                \State If the following conditions hold, for some $\tau^*$, then return 1:
                \begin{itemize}
                    \item $\GVf(ipk,nk^*) \land \textit{UVf}(nk^*, m^*,\sigma^*)$;
                    \item $(i^*, \tau^*) \in L_h \land i^* \notin L_{sk} \land (i^*,m^*,nk^*,*) \notin L_\sigma$;
                    \item $Judge(nk^*,i^*,ipk,\Pi^*) \lor 
 \textit{Trace}(ipk, \tau^*,nk^*)$.
                \end{itemize}
                \State return 0
            \end{algorithmic}
    \paragraph{$Exp_{NGS,\adversary}^{OS}(\lambda)$}
            \begin{algorithmic}
                \State $(isk,ipk) =IKg(1^\lambda); (osk, opk) = OKg(1^\lambda)$
                \State $(nk^*, i_0^*,\Pi_0^*,i_1^*,\Pi_1) = \adversary^{\AddU,\CrptU,\SndToI,\USK,\RReg}(ipk,osk)$
                \State If the following conditions hold, then return 1:
                \begin{itemize}
                    \item $\GVf(ipk,nk^*)$;
                    \item $\textit{Judge}(nk^*,i_0^*,ipk,\Pi_0^*) \land
                    \textit{Judge}(nk^*,i_1^*, ipk,\Pi_1^*);$
                    \item $i_0^* \neq i_1^*$.
                \end{itemize}
                \State return 0
            \end{algorithmic}
                     \paragraph{$Exp_{NGS,\adversary}^{OC}(\lambda)$}
            \begin{algorithmic}
                \State $(isk,ipk) = IKg(1^\lambda);
                (osk, opk) = OKg(1^\lambda)$
                \State $(nk^*,i^*,\Pi^*) = \adversary^{\AddU,\CrptU,\SndToI,\USK,\RReg}(ipk,osk)$
                \State If the following conditions hold, for some $(i, \tau)$, then return 1:
                \begin{itemize}
                    \item $\GVf(ipk,nk^*)$;
                    \item $(i,\tau) \in L_h \land i\neq i^*$;
                    \item $\textit{Trace}(ipk, \tau,nk^*)\land 
 Judge(nk^*,i^*,ipk,\Pi^*).$ 
                \end{itemize}
                \State return 0
            \end{algorithmic}
   \paragraph{$Exp_{NGS,\adversary}^{Anon-b}(\lambda)$}
            \begin{algorithmic}
                \State
                $(isk,ipk) = IKg(1^\lambda); (osk, opk) = OKg(1^\lambda)$
                \State $b' = \adversary^{{\SndToU},{USK},{\WReg},{\Sig},{\Trace}, Ch_b}(opk,isk)$
                \State return $b'$
            \end{algorithmic}
        \end{algorithm}
        \caption{Traceability, Non-frameability, Optimal opening soundness, Opening coherence and Anonymity (from top to bottom)  experiments for NGS }
        \label{fig:trace}
    \end{figure}

    \subsubsection{Non-frameability}
    The issuer and opener are both controlled by $\adversary$, so he receives $osk$ and $isk$. The $\CrptU,\WReg$, and $Open$ oracles are therefore not needed.
    He can use the $\SndToU$ oracle to add a new honest user to the group and the $\USK$ oracle except for the target identity $i*$ he outputs.
      Finally, $\adversary$ can query the $Sig$ oracle for nicknames already traced by the specified user via the $\Trace$ oracle, also provided.
      
    The goal of such an adversary $\adversary =\adversary^{\SndToU,\USK,\Trace,Sig}$ is to produce a forgery $(nk^*,\sigma^*)$ on a message $m^*$ passing the $\GVf$ and $\UVf$ verifications, such that either an honest user traces $nk^*$, or the target identity $i^*$ and proof $\Pi^*$ additionally provided are accepted by the $Judge$ algorithm.
    The non-frameability advantage $Adv_{NGS,\adversary}^{Nf}(\lambda)$  is defined here as for traceability.

    \begin{definition}
        A NGS scheme is \emph{non-frameable} if $Adv_{NGS,\adversary}^{Nf}(\lambda)$ is negligible, for a given $\lambda$ and any $\adversary$.
    \end{definition}

    \subsubsection{Optimal opening soundness}
    \label{secprop:OS}
        
    Here, $\adversary$ can corrupt all entities, including the opener and all users, but not the issuer.
    He can use the $\AddU,\CrptU,\SndToI,\USK$ and $\RReg$ oracles.
    The adversary wins if he produces a nickname $nk$ with two opening users and proofs $(i_0,\Pi_0)$ and $(i_1,\Pi_1)$, where $i_0 \neq i_1$, both accepted by the $Judge$ algorithm\footnote{As for traceability, a similar security property could be defined for $\textit{Trace}$, and proven along the same lines as here.}.
    The optimal-opening-soundness advantage $Adv_{NGS,\adversary}^{OS}(\lambda)$  for  $\adversary = \adversary^{\AddU,\CrptU,\SndToI,\USK,\RReg}$ is defined as for traceability.
    
    \begin{definition}
        A NGS scheme is \emph{optimally opening sound} if $Adv_{NGS,\adversary}^{OS}(\lambda)$ is negligible for a given $\lambda$ and any $\adversary$.
    \end{definition}
\subsubsection{Opening coherence}
 
In this experiment, the issuer is honest while the opener can be malicious. The goal of the adversary is to produce a nickname $nk^*$ that an honest user can trace, and an opening proof that points to another user. It is given the same oracles as the optimal opening soundness experiment.
The opening-coherence advantage $Adv_{NGS,\adversary}^{OC}(\lambda)$  for any adversary  $\adversary = \adversary^{\AddU,\CrptU,\SndToI,\USK,\RReg}$ is defined as for traceability.
        \begin{definition}
        A NGS scheme is \emph{opening coherent} if $Adv_{NGS,\adversary}^{OC}(\lambda)$ is negligible, for a given $\lambda$ and any $\adversary$.
    \end{definition}

    \subsubsection{Anonymity}
     We provide a restricted access to the $\Trace$ and $Sig$ oracles. The former traces nicknames for non-target users; the latter only signs when the provided nickname has been traced by the specified user in the former oracle.
    In this experiment, the issuer is corrupted, i.e., the adversary $\mathcal{A}$ is given the issuer's key, but not the opener's one.
    He can  use the $\SndToU$ oracle to add an honest user to the group and  write to the registry with the $\WReg$ oracle.
    He has also  access to the $Ch_b$ oracle, providing a way, given two honest users and a message, to receive a NGS-signed message and nickname from one of the two; note that $Ch_b$ simulates an honest sender that calls the $\textit{Nick}$ function.
    He  can also call the $\USK$ oracle to get the secret keys of all users, except the ones that are challenged in the $Ch_b$ oracles~\cite{boneh2004group} (for which the oracles check that their secret keys have not been exposed). Finally, he can get a nickname $nk$ traced by a specified user using the $\Trace$ oracle, and call the $\Sig$ oracle for this user, nickname $nk$ and a message $m$ of his choice.


    We define the advantage in the anonymity experiment described in Figure~\ref{fig:trace} for any polynomial time adversary $\adversary = \adversary^{{\SndToU},{\USK},{\WReg},{\Trace},{\Sig},
    Ch_b}$ as
    $Adv_{NGS,\adversary}^{Anon}(\lambda)=|\textnormal{Pr}[Exp^{Anon-0}_{NGS,\adversary}(\lambda)=1]-\textnormal{Pr}[Exp^{Anon-1}_{NGS,\adversary}(\lambda)=1]|.$

    \begin{definition}
        A NGS scheme is \emph{anonymous} if $Adv_{NGS,\adversary}^{Anon}(\lambda)$ is negligible for a given $\lambda$ and any $\adversary$.
    \end{definition}

    \section{Construction}\label{sec:construction}
    The NGS construction we propose is based on the SEP paradigm~\cite{camenisch2002signature}.
     These schemes have led to efficient GS from randomizable signatures such as the Pointcheval-Sanders (PS) ones~\cite{shortrandomizable} and, in particular, \cite{kim2021practical}, compatible with our application domain (see, e.g., NickHat, in Appendix~\ref{sec:application}). 
    During the NGS joining phase, the ``issuer'' computes a PS signature on the user $i$'s committed key, and saves it (resp., the ElGamal encryption \cite{elgamal1985public} of the user's trapdoor for the opener) in $\mathbf{mpk}$ (resp., $\mathbf{reg}$).
    Anyone can then generate a new nickname $nk$ for this new member by randomizing $\mathbf{mpk}[i]$, and the $\GVf$ verification function reduces to checking the validity of PS signatures.
       Using his trapdoor, member $i$ can trace this new nickname and prove possession of it by performing a signature proof of knowledge on a message of his choice. The opener decrypts all trapdoors to check which user's equivalence class identifies a member from a given nickname. 

    \paragraph{Types and variables}
    Let $\mathbb{G}_1, \mathbb{G}_2 $, and $\mathbb{G}_T$ be cyclic groups, with $ e: \mathbb{G}_1 \times \mathbb{G}_2 \xrightarrow{} \mathbb{G}_T$, a pairing map, and $g$, resp. $\hat{g}$, a generator of $\mathbb{G}_1$, resp. $\mathbb{G}_2$.
    Master public keys are arbitrary representations of users' equivalence classes and, therefore, of the same nature as nicknames; they belong to $\mathbb{G}_1^3$, and thus $l=3$ and $\mathbb{G} = \mathbb{G}_1$ in $\mathcal{R}$. 

    A join request $reqU$ consists of a tuple $(f,w,\tau',\pi_J,\sigma_{DS})$ with $(f,w) \in \mathbb{G}_1^2$, $\tau'\in \mathbb{G}_2^2$ (the El-Gamal encryption of a trapdoor $\tau$), $\pi_J$, the proof of knowledge for $PK_J$ computed during the $Join$ operation, and $\sigma_{DS}$, a $DS$ signature, for some Digital Signature scheme $DS$ parametrized over the security parameter $\lambda$.

    A registration information $reg$ consists of a tuple $(f,\tau', \rho, \sigma_{DS})$, with $f\in \mathbb{G}_1$, $\tau' \in \mathbb{G}_2^2$, $\rho \in \mathbb{G}_T$ and $\sigma_{DS}$, a DS signature.

    Finally, let $H:\mathbb{G}_1 \rightarrow \mathbb{G}_1$ be a hash function.

    \paragraph{Functions}
    Our scheme defines the functions of Section~\ref{sec:ngs} in three parts as follows.

    \begin{itemize}
        \item $IKg :1^\lambda \rightarrow (isk,ipk)$.
        Select $(x,y) \leftarrow_\$ \mathbb{Z}_p^2$, and compute $\hat{X}=\hat{g}^x$ and $\hat{Y}=\hat{g}^y$.
        Return $((x,y),(\hat{X},\hat{Y})).$
        \item $OKg : 1^\lambda \rightarrow (osk,opk)$. Select $z \leftarrow_\$ \mathbb{Z}_p$.  Return $(z,\hat{g}^z)$.
        \item $UKg :(i,1^{\lambda}) \rightarrow (usk,upk)$.
        Let $(usk, upk) = DS.KeyGen(1^\lambda)$. Set $\textbf{upk}[i] = upk$. Return $(usk, upk)$.
        \item $Join : (usk, opk) \rightarrow (msk, \tau, reqU) $.
        \begin{enumerate}[leftmargin=1.3\parindent]
            \item  Select $(\alpha, s) \leftarrow_\$ \mathbb{Z}_p^2$, and compute $f=g^\alpha, u=H(f)$ and $w=u^\alpha$.
            Let $\hat{Z}=opk$.
            Then, compute $\tau=\hat{g}^{\alpha}$ and its El-Gamal encryption $\tau'=(\hat{S},\hat{f'})$, with some $s$ such that $\hat{S}=\hat{g}^{s}$ and $ \hat{f}'=\tau \cdot \hat{Z}^{s}$.
            \item Generate a proof $\pi_J$ for all the above definitions, with $\pi_J=PK_J\{(\alpha,s): f=g^\alpha \land w=u^\alpha  \land   \hat{S}=\hat{g}^{s} \land
            \hat{f}'=\hat{g}^\alpha \cdot \hat{Z}^{s} \}$.
            \item Finally, let $\sigma_{DS}=DS.Sig(usk,\rho)$, where $\rho=e(f,\hat{g})$, $msk=\alpha$ and $reqU=(f,w,\tau',\pi_J,\sigma_{DS})$. Then, return  $(msk, \tau,reqU)$.
        \end{enumerate}

        \item $Iss: (i,isk,reqU, opk) \rightarrow ()$ or $\perp$.
        Let $(f,w,\tau',\pi_J,\sigma_{DS})=reqU$ and $(x,y)=isk$.
        Compute $u=H(f)$ and $\rho=e(f,\hat{g})$ and check the following (return $\perp$ if this fails):
            (1) $f$ did not appear  previously;
            (2) $\pi_J$ is valid for $\hat{Z}=opk$;
            (3) $\sigma_{DS}$ is valid on $\rho$ under $\mathbf{upk}[i]$.
        Then compute $v=u^x \cdot w^y$ and set $\mathbf{reg}[i] = (f,\tau',\rho,\sigma_{DS})$ and $\mathbf{mpk}[i] = (u,v,w)$.
        Return $()$.
        ~\\
        \item $Nick : (mpk, r) \rightarrow nk$. Let $(u,v,w)=mpk$.
        Then, 
        let $nk=(u^r,v^r,w^r)$. Return $nk$.
        \item $Trace : (ipk, \tau,nk) \rightarrow b$. Let $(u,v,w)=nk$ and $b=(e(u,\tau)= e(w,\hat{g})) \land \GVf(ipk,nk)$. Return $b$.

        \item $\GVf~: (ipk,nk) \rightarrow b$. Let $(u,v,w)=nk$ and $(\hat{X},\hat{Y})=ipk$. Compute and return  $b = (e(v,\hat{g})=e(u,\hat{X})\cdot e(w,\hat{Y}))$.
        
        \item $Open : (osk,nk) \rightarrow (i,\Pi) \text{ or} \perp$. Let $(u,v,w)=nk$ and $z=osk$.
        \begin{enumerate}[leftmargin=1.3\parindent]
            \item For each $reg=(f,\tau',\rho,\sigma_{DS}) \in \mathbf{reg}$:
            \begin{enumerate}
                \item let $(\hat{S},\hat{f}')=\tau'$, and get the user trapdoor $\tau$ as  $\tau=\hat{f'}\cdot \hat{S}^{-z}$;
                \item check if ($e(u,\tau)=e(w,\hat{g}) \land \rho=e(g,\tau))$.
            \end{enumerate}
            \item If Step 1 fails for all $reg$, then output $\perp$, and, otherwise, let $i$ the index of the (unique) $reg$ that succeeds;
            \item Compute $\pi_O=PK_O\{(\tau):e(w,\hat{g})=e(u,\tau) \land \rho=e(g,\tau)\}$;
            \item With $\Pi=(\rho,\sigma_{DS}, \pi_O)$, return $(i,\Pi)$.
        \end{enumerate}

        \item  $Judge : (nk,i,ipk,\Pi) \rightarrow b$.
        First, let $(\rho,\sigma_{DS}, \pi_O)=\Pi$ and  $(u,v,w)=nk$; get $upk=\mathbf{upk}[i]$.
        Then, check:
            (1) the validity of $\pi_O$;
            (2) whether $DS.\textit{Vf}(upk,\rho,\sigma_{DS})=1$;
            (3) the value of $\GVf(ipk,nk)$.
        Set $b$ to $true$, if the three conditions hold, and to \textit{false}, otherwise, and return $b$.
        ~\\
                \item $Sign : (nk,msk,m) \rightarrow \sigma$.
        Let $(u,v,w)=nk$ and $\alpha=msk$. Compute and return $\sigma = SPK_S\{(\alpha): w=u^\alpha\}(m)$.
        \item $\textit{UVf}~: (nk,m,\sigma) \rightarrow b$. Let $(u,v,w)=nk$ and return $b  = true$, if the signature proof of knowledge $\sigma$ is valid with respect to $(u,w)$, and $m$, \textit{false} otherwise.

    \end{itemize}

   These functions represent a correct construction of the NGS
interface of Section~\ref{sec:interface}, which can be proven by case analysis (see Appendix~\ref{sec:proofs}).

    \paragraph{Security analysis}
Assuming that the SPKs used above are simulation-sound extractable NIZKs and that $H$ is modeled as a random oracle, then one can prove (see Appendix~\ref{sec:proofs}), following the definitions in Section~\ref{sec:security}, that this NGS construction is non-frameable (under the EUF-CMA of the digital signature scheme and the Symmetric Discrete Logarithm assumption), optimally opening sound, opening coherent, traceable (under the Modified GPS assumption~\cite{kim2021practical}) and anonymous (under the Symmetric External Diffie-Hellman assumption, SXDH).

\section{Conclusion and Future work}\label{sec:conclusion}
We introduced Nicknames for Group Signatures (NGS), a new signing scheme that seamlessly combines anonymous transfers from signatures with flexible public keys and auditability from group signatures. 
We formally defined the NGS security model and described an efficient construction, which we proved secure in the Random Oracle Model. 
We developed a Rust implementation of NGS used in NickHat, an auditable, compliant solution for any ERC20-based token in Ethereum. 

    Our NGS construction focuses on efficiency for our Ethereum application. 
    Regarding future work, one might want to enforce a stronger traceability property with the adversary only asked for a nickname that passes the group verification function and cannot be opened. 
Additionally, stronger anonymity with full IND-CCA security would also be desirable, with an $Open$ oracle fully available. 
To address decentralization requirements, one could also aim for threshold issuance and opening, as done in GS~\cite{camenisch2020short}. 
    The opening could also be made more efficient by identifying directly the suspect without requiring a linear scanning~\cite{kim2023practical}.
    Finally, a user revocation feature needs to be added, e.g., in the line of~\cite{LiPe12}.
    

\bibliographystyle{abbrv}
\bibliography{base}

\appendix


    \section{NickHat}\label{sec:application}
    
    
    We introduce NickHat, an NGS-based permissioned token-exchange system that supports auditing and provides anonymity on Ethereum. 
    It requires an on-chain verifier, made possible by the available pairing-check precompile of the BN254 curve \cite{barreto2005pairing}.
    As in~\cite{hopwood2016zcash}, NickHat supports two address types: the transparent ones, which are Ethereum standard addresses, and the private ones, related to nicknames. Typically, one can transfer tokens from a transparent address to a private one (to enter NickHat), from a private address to another private one (within NickHat), and finally, from a private address to a transparent one (thus transferring tokens out of NickHat).  Although NickHat could work with any type of Ethereum currency, we focus, in this prototype, on tokens that respect the ERC20 standard. 

    Roles in NickHat are close to NGS's ones.
    We keep the issuer, verifier, and user denominations; the NGS' opener and judge are renamed ``supervisor'' and ``(external) auditor'', respectively. We add a ``relayer'' role that takes specific transactions (following the ERC-2771 standard) as inputs and sends them to the blockchain and an ``operator'' that manages the system.
    We describe below the main operations of NickHat.

\begin{description}[leftmargin=0.2cm]
  \item[Setup] First, an issuer and a supervisor are chosen by the operator; they run the NGS $IKg$ and $OKg$ functions, respectively, to obtain their keys. 
    In the blockchain environment, the operator deploys four smart contracts: 
    \begin{itemize}
        \item the verifier contract, which embeds the NGS $\UVf$ and $\GVf$ functions for the chosen issuer;
        \item the key-registry contract, which adds master public keys and displays them; 
        \item the NickHat contract, which provides the \textit{deposit}, \textit{transfer} and \textit{withdraw} functions (see below), keeping a \textit{balances} array up-to-date; 
        \item  the forwarder contract, with the \textit{execute} function that redirects requests to their specified destination after verifying the NGS signatures.
    \end{itemize}
  \item[User registration] When a user wants to join the issuer-managed group and use NickHat, he runs the NGS group-joining algorithm with the issuer and obtains, if successful, his master public key. The issuer gets the user's encrypted trapdoor. Finally, the operator saves the new master public key by sending a transaction to the key-registry contract, which makes it publicly readable.
  \item[Deposit] Assume user $i$ already holds $v$ units of a deployed ERC20 token.
    User $i$ wants to transfer these tokens to group member $j$ without disclosing the latter's identity. To do this, user $i$ first sends a signed transaction to the ERC20 token that allows the NickHat contract to dispose of his $v$ tokens. 
    Next, user $i$ gets $j$'s master public key $mpk$ from the key-registry contract and computes a nickname $nk = Nick(mpk,r)$ for $j$ and some random coin $r$ via the NGS algorithm. He then signs a transaction to deposit $v$ tokens to $nk$ and sends it to
    the \textit{deposit} function of the NickHat contract. This contract spends the ERC20 token allowance and transfers $i$'s tokens to itself, acting as an escrow. The NickHat \textit{balances} array is then updated by setting $v$ tokens for the specified nickname $nk$ and token address. Finally, the NickHat contract emits an Announcement event with the nickname $nk$. 
    \item[Detection]
           All group members listen to the Announcement events from the blockchain and call the \textit{Trace} function to check if some of the new nicknames mentioned there belong to them. If so, they know that some tokens have been kept in escrow for them, which they can use as they wish.
    \item[Transfer] Assume that group member $i$ has $v$ escrowed tokens linked to one of his nicknames, $nk$, and wants to sends these  tokens to group member $j$. He first gets  $j$'s master public key \textit{mpk'} from the key-registry contract and calls the NGS \textit{Nick} function to obtain a new nickname $nk' = Nick(mpk',r')$ for some random coin $r'$. Then, he prepares a request $\rho$ asking to send $v$ tokens from $nk$ to $nk'$ and signs $\rho$ with the NGS $Sign$ function on the hash value of $\rho$. A relayer is then handed $\rho$ and wraps it to a meta-transaction (ERC-2771) that he signs and sends to the \textit{execute} function of the forwarder contract. The forwarder contract first verifies the NGS signature by calling the \textit{UVf} function of the verifier contract and redirects the call to the \textit{transfer} function of the NickHat contract. This contract then updates \textit{balances} (so that $balances[nk'] = v$ and $balances[nk] = 0$) and emits an Announcement event regarding $nk'$.
    \item[Withdraw] User $i$ listens to Announcement events and learns that some $nk$ mentioned is his. To withdraw his escrowed tokens to an Ethereum address $a$, he must proceed along the following way. 
    A request $\rho$ is first prepared, stating that the owner of $nk$ sends $v$ tokens to $a$, and is then handed to a relayer.
    The relayer wraps the request into a transaction and sends it to the \textit{execute} function of the forwarder contract. This contract (1) verifies the signature by calling \textit{UVf},  of the verifier contract, (2) redirects the call to the NickHat \textit{Withdraw} function, which subtracts $v$ tokens from $balances[nk]$ (if possible), and (3) asks the ERC20 token contract to transfer $v$ tokens from the NickHat address to $a$.
    \item[Audit] The supervisor can inspect any nickname announced in blockchain Announcement events. He can retrieve the identity of the user behind a nickname by calling the first part of the NGS \textit{Open} algorithm. In the case of an audit request on a particular nickname $nk$, the supervisor can prove to the external auditor that a certain group member is indeed behind $nk$. The external auditor can run the \textit{Judge} function to be convinced. 
\end{description}

\paragraph{Performance}
    In terms of computation, the Transfer protocol requires the randomization of the recipient's nickname, i.e., 3 exponentiations in $\mathbb{G}_1$, and the signature proof of knowledge of the sender in $\mathbb{G}_1$, totaling 4 exponentiations. Note that this computation is performed off-chain. 
    The verification protocol, on-chain, consists of the recipient's nickname group verification, i.e., 3 pairing computations, and the sender proof of knowledge, i.e., 2 exponentiations in $\mathbb{G}_1$. 
    
    To give an idea of the monetary costs induced by the use of NGS, we provide in Table~\ref{tab:eth_tx_costs_operations} the cost of each NickHat-related operation. 
    
    \paragraph{Extensions} In addition to NickHat, we tested other financial use cases for NGS, in particular in a delivery-vs-payment (DVP) setting, where securities can be atomically exchanged for cash. As explained above, the NickHat contract already handles ERC20 tokens, and it is sufficient for DVP applications to extend it with hashed time-lock contract (HTLC)~\cite{poon2016bitcoin} functions. 
    After approving the operation, users can lock their NickHat tokens in the HTLC contract. They can then claim their entitled tokens, or request a refund.

\begin{table}
\centering

\begin{tabular}{|l|r|r|r|}
\hline
\textbf{Operation} & \textbf{Consumption (Gas)} & \textbf{Cost ($10^{-4}$ ETH)} & \textbf{Cost (USD)} \\
\hline
Insert   & 39,230   & \( 0.3962 \) & \$0.10 \\
Deposit  & 263,610  & \( 2.668  \) & \$0.69 \\
Transfer & 292,480  & \( 2.954  \) & \$0.76 \\\hline
Approve  & 86,450   & \( 0.8711 \) & \$0.23 \\
Lock     & 248,980  & \( 2.515  \) & \$0.65 \\
Claim    & 369,830  & \( 3.735  \) & \$0.96 \\
Refund   & 308,570  & \( 3.117  \) & \$0.80 \\
\hline
\end{tabular}
\caption{Transaction costs for NickHat-related operations (the gas price was 1.01 Gwei and 1 ETH was worth \$2,577, on March 7, 2025). The Insert operation consists of publishing one master public key. The Deposit and Transfer operations correspond to anonymously depositing and transfering existing tokens into the NickHat contract. The Approve, Lock, Claim and Refund operations correspond to the DVP use case mentioned in the text, where users have to lock their tokens and can claim or refund their dues. }
\label{tab:eth_tx_costs_operations}
\end{table}

\section{Proofs}\label{sec:proofs}
    \subsection{Traceability proof}\label{proof:trace}
    $\challenger$ is given an instance of the Modified GPS problem, i.e., a pair $(\hat{X},\hat{Y})$ such that their respective discrete logarithms $(x,y)$ are unknown to him, and the two oracles $\mathcal{O}_0^{MGPS}$ and $\mathcal{O}_1^{MGPS}$.
    $\challenger$ acts as an honest issuer and maintains the lists ${L}_0^{MGPS}$ and ${L}_1^{MGPS}$ for  $\mathcal{O}_0^{MGPS}$ and $\mathcal{O}_1^{MGPS}$, respectively, storing input/output values.
    Also he can use $\mathcal{E}$, the extractor of $PK_S$.

    $\challenger$ sets $ipk=(\hat{X},\hat{Y})$ and samples $osk=(z_0,z_1) \leftarrow_\$ \mathbb{Z}_p^2$. It then answers the different oracles queries as follows, assuming that $\adversary$ has access to $ipk$ and $osk$ ($\CrptU, \USK$ and $\RReg$ operate as already specified, and $L_H$ is an initially empty mapping for hashed values).
    \begin{itemize}
        \item Hash. Given input $n$ such that $(n,u) \in {L}_H$ for some $u$, return $u$.
        Otherwise, if $n\in \mathbb{G}_1$, $\challenger$ calls $\mathcal{O}_0^{MGPS}$ to get some $u\in \mathbb{G}_1$ and adds $(n,u)$ to $L_H$ and $u$ to ${L}_0^{MGPS}$.
        Otherwise, $\challenger$ samples $u \leftarrow_\$ \mathbb{G}_1$ and add $(n,u)$ into $L_H$.
        Finally, $u$ is returned.
        \item AddU. $\challenger$ follows the $Addu$ protocol to add the user $i$. However, since it does not know $isk$,
        $v$ is obtained by calling $\mathcal{O}_1^{MGPS}$ ($L_1^{MGPS}$ gets thus updated) with $(g,u,f,w)$ as input computed following the protocol.
        Finally, $(i, \tau)$ is added to ${L}_h$.
        \item SndToI: For a queried identity $i$ such that $(i,cont) \in {L}_c$ (i.e., $i$ comes from $\CrptU$; otherwise $\perp$ is returned), $\challenger$ also receives its join request $r = (f,w,\tau',\pi_J,\sigma_{DS})$.
        $\challenger$ checks that $f$ is unique (i.e., not in $\mathbf{reg}$), that $\sigma_{DS}$ is a valid signature for the message $\rho = e(f,\hat{g})$ and public key $\mathbf{upk}[i]$, and that $\pi_J$ is also valid.
        $\challenger$ then computes $u = Hash(f)$, therefore obtained from $\mathcal{O}_0^{MGPS}$.
        As in the $\CrptU$ oracle, $\challenger$ does not know $isk$ and calls $\mathcal{O}_1^{MGPS}$ with $(g,u,f,w)$ as input to get $v=u^x \cdot w^y$;
        ${L}_1^{MPGS}$ is, similarly, updated after the call to $\mathcal{O}_1^{MGPS}$. Then,
        $\mathbf{mpk}[i]$ is set to $(u,v,w)$, and $\mathbf{reg}[i]$, to $(i,\tau',\rho,\sigma_{DS})$.
        ${L}_c$ is updated with $(i,accept)$.

    \end{itemize}

    At the end, $\adversary$ outputs its forgery $nk^*=(u^*, v^*, w^*)$, $\sigma^*$ and $m^*$ with $\sigma^*$ valid on the message $m^*$.
    Now, as $\GVf(nk^*)$ and $\sigma^*$ should be valid, as stated in the experiment, two cases are possible.
    \begin{itemize}
        \item  $Open(osk,nk) = \perp$.
        We make the contradiction explicit.
        For each $reg=(i,\tau',\rho,\sigma_{DS}) \in \mathbf{reg}$, $\challenger$ decrypts $\tau'$ to get $\tau$.
        In this case, for each $reg$, at least one of the two pairing check fails, i.e., $e(u^*,\tau)\neq e(w^*,\hat{g})$ or $\rho \neq e(g,\tau)$.
        The second inequality would break the simulation soundness of $PK_J$ and is therefore excluded.

        Thus, $w^* \neq {u^*}^{\alpha}$ for each user's $\alpha$ such that $\tau = \hat{g}^\alpha$.
        Therefore, $\alpha^*=\log_{u^*}w^*$ is new, i.e., not stored in $\mathbf{reg}$.
        But, then, $\challenger$ could use $\extractor$ of $PK_S$ on $\sigma^*$ to extract this $\alpha^*$ and solve the MGPS problem with $((u^*,v^*),\alpha^*)$, a contradiction.
        \item There exist $i^*$ and $\Pi^*$ with $(i^*,\Pi^*) = Open(osk,nk)$ and $ Judge(nk^*,ipk,i^*,\Pi^*)$ is false.
        As the $Open$ algorithm is run honestly, this case indicates that the $\tau^*$ encrypted in $\mathbf{reg}[i^*]$ satisfies both $e(u^*,\tau^*)=e(w^*,\hat{g})$ and $\rho=e(g,\tau^*)$.
        Thus, since the proof $\pi_O$  holds with these values and $\sigma_{DS}$ on $\rho$ has already been verified in the $\SndToI$ oracle, $Judge$ should be true, a contradiction.
    \end{itemize}

    \subsection{Non-frameability proof}\label{proof:nf}

    Let $(g,\hat{g},D, \hat{D})$, with $D = g^d$ and $\hat{D}=\hat{g}^d$, be an instance of the SDL (Symmetric Discrete-Logarithm) problem for some unknown $d$, and let $\simulator$ be a zero-knowledge simulator and $\extractor$, a knowledge extractor for PKs.
    Let $q$ be the number of queries to the $\SndToU$ oracle to be made by $\adversary$. The challenger $\challenger$ picks a random $k \in \{1,..,q\}$ and hopes that the target user $i^*$ corresponds to the identity of the $k^{th}$ query.

    We describe below how the challenger $\challenger$ responds to the relevant oracles. It uses a caching table $L_H$, initialized with the tuple $(D, \delta)$, where $\delta \leftarrow_\$ \mathbb{Z}_p$.
    \begin{itemize}
        \item Hash: For the queried input $n$, if $(n,\delta) \in {L}_H$, $\challenger$ returns $g^\delta$.
        Otherwise, $\challenger$ samples $\delta \leftarrow_\$ \mathbb{Z}_p$,  stores $(n,\delta)$ in ${L}_H$ and returns $g^\delta$.
        \item SndToU: For the queried identity $i$, if $i$ hasn't been queried before, $\challenger$ runs $UKg$ to get its $(usk,upk)$; otherwise, $(usk, upk) = (\mathbf{usk}[i], \mathbf{upk}[i])$.
        Then, if this is not the $k^{th}$ query, $\challenger$ completes the original $\SndToU$.
        Otherwise,  $\challenger$ sets $i^* = i$, samples $(s_0,s_1) \leftarrow_\$ \mathbb{Z}_p^2$ and gets the value $(D,\delta) \in {L}_H$.
        With $(\hat{Z}= opk$, it sets $\mathbf{reqU}[i^*] = reqU $ with $reqU = (D,D^{\delta}, \tau', \pi, \sigma_{DS})$, where
        $\sigma_{DS} = DS.Sig(\mathbf{usk}[i^*], \rho)$, $\tau'$ is the encryption pair $(\hat{g}^{s}, \hat{D}\hat{Z}^{s})$ of $\tau=\hat{D}$, $\rho=e(D,\hat{g})$ and $\pi$ is a simulated proof of knowledge of the unknown $d$ and of $(s_0, s_1)$, produced by $\simulator$.
        Finally, $\challenger$  returns $reqU$ and adds $(i,\tau)$ to $L_h$.
        \item USK: for a queried identity $i$, if $ i \notin {L}_h$ or $i=i^*$, i.e., $\SndToU$ has been called at least $k$ times, then $\challenger$ aborts.
        Otherwise, it sends $(\mathbf{usk}[i], \mathbf{msk}[i])$ to $\adversary$.
        \item Trace: for a queried identity $i$ and nickname $nk$, $\challenger$ follows the oracle protocol.  
        \item Sig: for a queried identity $i$, nickname $nk$, and message $m$,
        $\challenger$ follows the protocol except that he simulates the signature proof of knowledge  on the message $m$ for $nk$ using the simulator $\simulator$ and returns it.
    \end{itemize}
    At the end of the experiment, $\adversary$ outputs a signing forgery $(nk^*,\sigma^*)$ on a message $m^*$ with $(u^*,v^*,w^*)=nk^*$ along with the opening result $(i,\Pi^*)$, with $\Pi^* =(\rho^*,\sigma^*_{DS},\pi^*_O)$.
    If $i \not= i^*$, $\challenger$ aborts.
    Otherwise, let $(f_k, w_k, \tau'_k, \pi_k,\sigma_{k}) = \mathbf{reqU}[i^*]$.
    $\challenger$ decrypts, with the opening key $opk$, $\tau'_{k}$ to get $\tau_{k}$.
    
    When the forgery is accepted by the $Judge$ algorithm, $\pi_O^*$ is valid and $\sigma_{DS}^*$ is a valid signature of $\rho^*$ for $\mathbf{upk}[i^*]$.
    We consider the 3 possible cases of forgery, where $\rho_k$ is the pairing signed in $\sigma_k$.
    \begin{itemize}
        \item $\rho_k \neq \rho^*$. Since $\text{$\textit{DS.Vf}$}(\mathbf{upk}[i^*], \rho^*, \sigma^*_{DS})$, this indicates that the signature $\sigma^*_{DS}$ on the message $\rho^*$, which the honest user $i^*$ did not sign, is valid.
        This breaks the unforgeability of $DS$.
        \item $e(w^*,\hat{g})=e(u^*,\tau)$, for some $\tau$ different from $\tau_k$. This indicates that the statement required in the $Open$ function is false, thus breaking the simulation soundness of $PK_O$.
        Indeed, since $\rho_k =\rho^*$, one has $\rho_k = e(g^d, \hat{g}) =  e(g, \hat{g}^d) = \rho^* =  e(g, \tau^*)$, for some $\tau^*$ referenced in $\pi^*_O$. Since $e$ is injective in each of its dimensions, one has $\tau^* = \hat{g}^d = \tau_k$ and, from $\pi^*_O$, $e(w^*,\hat{g})$ should be equal to $e(u^*,\tau_k)$.
        \item
        $e(w^*,\hat{g})=e(u^*,\tau_k)$. This last case yields $w^* = (u^*)^d$, and $\challenger$ can use the knowledge extractor $\extractor$ on $\sigma^*$ to get the witness $d$, i.e., $\log_{u^*}(w^*)$, for  the SDL challenge, thus solving it.
    \end{itemize}

Otherwise, if $Trace(ipk,\tau_k,nk^*)$ is true, this second case comes down to the last bullet point above, where we can extract the witness $d$ for the SDL challenge.
    
    Overall, if $\adversary$ succeeds in the NGS non-frameability experiment on our construction with probability $\epsilon=Adv_{NGS,\adversary}^{Nf}$, then we showed that
    $$\epsilon \leq q (Adv_{DS,\adversary}^{EUF}+ Adv_{PK_O,\adversary}^{SS}+2Adv_{\adversary}^{SDL}),$$
    \noindent with $Adv_{DS,\adversary}^{EUF}$, the advantage of the adversary in breaking the unforgeability of the DS signature, $Adv_{PK_O,\adversary}^{SS}$,  the advantage of $\adversary$ in breaking the simulation soundness of $PK_O$, and $Adv_{\adversary}^{SDL}$, the advantage of $\adversary$ in breaking the SDL problem.

    \subsection{Opening soundness proof}\label{proof:os}
    In the join protocol, since the issuer is honest, it therefore prevents two users from having the same $\alpha$ by checking in  $\SndToI$ and $\CrptU$ that the same first element of $reqU$, $f$, doesn't appear in previous or current joining sessions.
    By the soundness of $\pi_J$, the encrypted $\tau=\hat{g}^\alpha$ in $\mathbf{reg}[i]$ is also uniquely assigned to user $i$.

    Now, at the end of the experiment, $\adversary$ outputs $(nk^*,i_0^*,\Pi_0^*, i_1^*,\Pi_1^*)$ with $nk^*=(u,v,w)$ that successfully passes the $\GVf$ verification, and $\Pi_0^* =(\rho,\sigma_{DS}^*,\pi_O)$  and $\Pi_1^* $
    successfully verified by the $Judge$ algorithm for two distinct users, $i_0^*$ and $i_1^*$, respectively.
    We now show a contradiction.
    Let $w=u^\alpha$ for some $\alpha \in \mathbb{Z}_p$.
    Since $(i_0^*, \Pi_0^*)$ is accepted by the $Judge$ algorithm, $\rho=e(g,\tau_0)$ for some $\tau_0$ linked to $i_0^*$ by $\sigma_{DS}$.
    Then, by the soundness of $\pi_O$, it holds that $e(w,\hat{g})=e(u,\tau_0)$ for the same $\tau_0$.
    Thus, $\tau_0=\hat{g}^{\alpha}$, where $\alpha$ is the exponent that was uniquely assigned to user $i_0^*$ in the join protocol.
    But, with the same reasoning, $\alpha$ is the exponent that was uniquely assigned to user $i_1^*$ in the join protocol.
    But we stressed before that the join protocol prevents two users from having the same $\alpha$ exponent; this is therefore a contradiction.

    Overall, if $\adversary$ succeeds in the NGS opening soundness experiment on with above construction with probability $\epsilon=Adv_{NGS,\adversary}^{OS}$, then we showed that
    $$\epsilon \leq  Adv_{PK_O,\adversary}^{SS}+Adv_{PK_J,\adversary}^{SS},$$
    \noindent
    where  $Adv_{PK_J,\adversary}^{SS}$ is the advantage of $\adversary$ in breaking the simulation soundness of $PK_J$, and $Adv_{PK_O,\adversary}^{SS}$, the one of $PK_O$.

        \subsection{Opening coherence proof}\label{proof:oc}
    In the join protocol, since the issuer is honest, it therefore prevents two users from having the same $\alpha$ by checking in  $\SndToI$ and $\CrptU$ that the same first element of $reqU$, $f$, doesn't appear in previous or current joining sessions.

    Now, at the end of the experiment, $\adversary$ outputs $(nk^*,i^*,\Pi^*)$ with $nk^*=(u,v,w)$ that successfully passes the $\GVf$ verification but also is traced by some honest user $i$, $(i^*,\Pi^*)$
    successfully verified by the $Judge$ algorithm with $\Pi^* =(\rho^*,\sigma_{DS}^*,\pi_O)$ and $i\neq i^*$.
    We now show a contradiction.
    Let $w=u^\alpha$ for some $\alpha \in \mathbb{Z}_p$.
    Since $(i^*, \Pi^*)$ is accepted by the $Judge$ algorithm, $\rho=e(g,\tau^*)$ for some $\tau^*$ linked to $i^*$ by $\sigma_{DS}$.
    Then, by the soundness of $\pi_O$, it holds that $e(w,\hat{g})=e(u,\tau^*)$ for the same $\tau^*$.
    Thus, $\tau^*=\hat{g}^{\alpha}$, where $\alpha$ is the exponent that was uniquely assigned to user $i^*$ in the join protocol.
    But user $i$ is honest and the trapdoor $\tau$ of user $i$ is the one from the join protocol, thus binding this user to $\alpha$.

    But we stressed before that the join protocol prevents two users from having the same $\alpha$ exponent; this is therefore a contradiction.

    Overall, if $\adversary$ succeeds in the NGS opening coherence experiment on with above construction with probability $\epsilon=Adv_{NGS,\adversary}^{OC}$, then we showed that
    $$\epsilon \leq  Adv_{PK_O,\adversary}^{SS}+Adv_{PK_J,\adversary}^{SS}.$$

    \subsection{Anonymity proof}\label{proof:anonymity}
    For this proof, we chain experiments from $G_0$, the original experiment $Exp^{Anon-0}_{GNS,\adversary}$, to $G_F$, a final experiment where the secret key $\alpha$ of some user $i_0$ is can be replaced by any random key. We prove indistinguishability between these experiments.
    
    Note that the complete proof requires a subsequent sequence of hybrid experiments from $G_F$ to $Exp^{Anon-1}_{GNS,\adversary}$, this time, but we omit it as it can be deduced by reversing the first sequence we present.
    We describe below the experiments with their proofs of indistinguishability.
    \begin{itemize}
        \item  $G_0$ is the original security experiment $Exp^{Anon-0}_{GNS,\adversary}$  for the target identity $i_0$.
        \item $G_1$ is the same as $G_0$, except that $\challenger$ simulates all PKs without witnesses, assuming it  can use $\simulator$, the zero-knowledge simulator of PKs.
        By the zero-knowledge property of PKs, $G_1$ is indistinguishable from $G_0$.
        \item $G_2$ is the same as $G_1$, except that $\challenger$ aborts when $\adversary$ produces a proof on a false statement for the PKs.
        $G_2$ is indistinguishable from $G_1$ by the simulation soundness of the PKs.
        \item $G_3$ is the same as $G_2$, except that $\challenger$ guesses the target identities $i_0$ and $i_1$ that will be queried by the first call to $Ch_0$ oracle.
        Suppose there are $q$ queries sent to the $\SndToU$ oracle, $\challenger$ picks a random pair $(k_0,k_1) \leftarrow_\$ \{1,..,q\}^2$ and selects the $k_0$-th and $k_1$-th queries to the $\SndToU$ to register, from $L_h$, the $i_0$ and $i_1$ identities.
        If the guess is wrong, $\challenger$ aborts.
        This happens with probability at most $\frac{1}{q^2}$ and thus reduces the advantage by ${q^2}$.
        \item $G_4$ is the same as $G_3$, except that, in $\SndToU$, $\challenger$ replaces $\hat{f}'$, in the $\tau$ for $ \mathbf{reg}[i_0]$, by some random element from $\mathbb{G}_2$.
        By Lemma \ref{lemma:anon-g4}, $G_4$ is indistinguishable from $G_3$.
        \item $G_5$ is the same as $G_4$, except that, in $\SndToU$, $\challenger$ sends a $reqU$ request for user $i_0$ with $\log_uw\neq \log_gf$.
        By Lemma \ref{lemma:anon-g5}, $G_5$ is indistinguishable from $G_4$.
        \item $G_6$ is the same as $G_5$, except that in $\SndToU$, $\challenger$ sends a $reqU$ request for user $i_1$ with $\log_uw\neq \log_gf$.
        By Lemma \ref{lemma:anon-g6}, $G_6$ is indistinguishable from $G_5$.
        \item $G_F$ is the same as $G_6$, except that, for $(u,w)$ and $(u',w')$ sent in the $\SndToU$ oracle for users $i_0$ and $i_1$, respectively, we have $\log_uw=\log_{u'}w'$.
        $G_F$ is indistinguishable from $G_6$ under Lemma \ref{lemma:anon-gf}.
    \end{itemize}

    \begin{lemma}
        \label{lemma:anon-g4}
        Let ${H}$ be modeled as a random oracle.
        Then, $G_4$ and $G_3$ are indistinguishable under the SXDH assumption for ${\mathbb{G}_2}$.
    \end{lemma}

Given $(\hat{g}, \hat{A},\hat{B}, T)$, an instance of the SXDH assumption on $\mathbb{G}_2$, where $(\hat{A},\hat{B})= (\hat{g}^a,\hat{g}^b)$ for some unknown pair $(a, b)$ and $T$ is given. 
    First, $\challenger$ runs $IKg$ and sets $(isk, ipk)$ with its return value. Finally, $\challenger$ sets $opk=\hat{B}$ ($osk$ will not be needed). 
    \begin{itemize}
        \item SndToU: if  $i \neq i_0$ and $i \neq i_1$, $\challenger$ follows the default protocol, by calling $UKg$ and $Join$, except that the proof $\pi_J$ is simulated.
        Otherwise, $\challenger$ picks $\alpha \leftarrow_\$ \mathbb{Z}_p$ and simulates $\pi_J$ with $\simulator$ for
        $(f, w,\hat{S},\hat{f'}) =(g^\alpha, H(g^\alpha)^\alpha,\hat{A},\hat{g}^\alpha \cdot T)$.
        This tuple along with $(\pi_J, \sigma_{DS})$, where $\sigma_{DS}$ is a signature on $\rho=e(f,\hat{g})$, are used in the returned $reqU$.
        At the end, $\challenger$  sets $\mathbf{msk}[i]=\alpha$ and updates the list of honest users: ${L}_h={L}_h \cup \{(i,\hat{g}^\alpha)\}$.
        \item USK: for an identity $i\in \mathcal{L}_h$, $\challenger$ returns $(\mathbf{usk}[i],\mathbf{msk}[i])$ to $\adversary$ if $i \neq i_0^* \land  i \neq i_1^*$, aborts otherwise.
        \item Trace: for an identity $i$ such that  $i \neq i_0^* \land  i \neq i_1^*$, $\challenger$ executes the oracle algorithm, and aborts otherwise.
        \item Sig: $\challenger$ executes the oracle algorithm, except that the proof of knowledge of $\textbf{msk}[i]$ is simulated.

        %
        %
    \end{itemize}

    Therefore, $\challenger$ simulates $G_3$, if $T=\hat{g}^{ab}$, and $G_4$, otherwise. Indeed, by definition of $\pi_J$, one has $\hat{f'} = \hat{g}^\alpha \cdot opk^s$, where, here, $s = a$ and $opk = \hat{B}$; this yields $\hat{f'} = \hat{g}^\alpha \cdot \hat{g}^{ab}$, which is to be compared with $\hat{g}^\alpha \cdot T$.
    If $\adversary$ can distinguish the two experiments, $\challenger$ could then use it to break SXDH on ${\mathbb{G}_2}$.

    \begin{lemma}
        \label{lemma:anon-g5}
        Let $H$ be modeled as a random oracle.
        Then $G_5$ and $G_4$ are indistinguishable under the XDH$_{\mathbb{G}_1}$ assumption.
    \end{lemma}

    $\challenger$ is given $(g,A,B,T)$, with $A=g^a$ and $B=g^b$, an instance of the SXDH assumption on ${\mathbb{G}_1}$, and the zero-knowledge simulator $\simulator$ of the SPKs.
    $\challenger$ first initializes $L_H$ with $\{(B, \perp, A)\}$.

    \begin{itemize}
        \item Hash: for a queried input $n$, if some $(n, \delta, \zeta) \in L_H$,  $\challenger$ returns  $\zeta$.
        Else, $\challenger$ first samples $\delta \leftarrow_\$ \mathbb{Z}_p$ and returns $g^{\delta}$.
        He then updates ${L}_H={L}_H \cup \{(n,\delta,\zeta)\}$.
        \item SndToU: if $i\neq i_0^*$, $\challenger$ follows the protocol, except that the proof is simulated.
        Otherwise, $\challenger$ sets $(f,w, \hat{S},\hat{f}') = (B,T, \hat{g}^{s},\hat{R})$, with $s \leftarrow_\$ \mathbb{Z}_p$ and $\hat{R} \leftarrow_\$ \mathbb{G}_2$. Let $\pi_J$ be the simulated proof of knowledge of $b=\log_{g} B$ and $s$, and $\sigma_{DS}$, the $DS$ signature on $\rho=e(B,\hat{g})$.
        $\mathcal{C}$ stores $\mathbf{msk}[i_0]=\perp$ and updates ${L}_h={L}_h \cup \{(i_0, \perp)\}$.
        Finally, the request built using these variables as in $Join$ is returned.
        \item Trace: for an identity $i$ such that $i \neq i_0^* \land  i \neq i_1^*$, $\challenger$ executes the oracle algorithm, and aborts otherwise.
        \item Sig: $\challenger$ executes the oracle algorithm, except that the proof of knowledge of $\textbf{msk}[i]$ is simulated.
        
        \item Ch$_0$: Let $(x,y)=isk$. For $i=i_0$, $\challenger$ samples $r \leftarrow \mathbb{Z}_p$ and computes $v =  u^{x} \cdot w^{y}$, with $(u,w) = (g^{r},B^{ r})$; otherwise, the standard protocol is followed.
        $\challenger$ then simulates the proof of knowledge $\sigma$ for $SPK_S$ of $b=\log_{g} B$.
        Finally, $\mathcal{C}$ adds $(i_0,m,nk,\sigma)$ and $(i_1,m,nk,\sigma)$  to ${L}_{ch}$ and returns $(nk,\sigma)$, with $nk=(u,v,w)$.
    \end{itemize}
    Since $T$ has been introduced as $w$ for $i_0$, $\challenger$ simulates $G_4$ if $T=g^{ab}$, and $G_5$ otherwise.
    Indeed, if $T=g^{ab}$, we have $f=B,u=H(f)=A$ and $w=u^b=T$, thus implementing a proper join protocol. Note also that $Ch_0$ has been modified to ensure a proper behavior even if $\mathbf{mpk}[i_0]$ is defined with values given in $\SndToU$.

    \begin{lemma}
        \label{lemma:anon-g6}
        Let $H$ be modeled as a random oracle.
        Then $G_5$ and $G_6$ are indistinguishable under the XDH$_{\mathbb{G}_1}$ assumption.
    \end{lemma}
    The proof follows the same strategy as in Lemma~\ref{lemma:anon-g4}, since $G_5$ is indistinguishable from $G_4$, but for user $i_1$.

    \begin{lemma}
        \label{lemma:anon-gf}
        Let $H$ be modeled as a random oracle.
        Then $G_6$ and $G_F$ are indistinguishable under the XDH$_{\mathbb{G}_1}$ assumption.
    \end{lemma}

$\challenger$ is given $(g,A,B,T)$, with $A=g^a$ and $B=g^b$, an instance of the SXDH assumption on ${\mathbb{G}_1}$, and the zero-knowledge simulator $\simulator$ of the SPKs.


    \begin{itemize}
        \item Hash: for a queried input $n$, if $n$ has already been queried, $\challenger$ returns $\zeta
        $ for the corresponding $(n,\zeta)$ in ${L}_H$.
        Otherwise, $\challenger$ samples $\delta \leftarrow_\$ \mathbb{Z}_p$, updates ${L}_H={L}_H \cup \{(n, g^\delta)\}$, and returns $g^{\delta}$.
        
        \item SndToU: if $i\neq i_0 \land i\neq i_1$, $\challenger$ follows the protocol, except that the proofs are simulated.
        Otherwise, he first updates ${L}_h={L}_h \cup \{(i,\perp)\}$.
        Then, if $i=i_0$, $\challenger$ samples $\alpha_0 \leftarrow_\$ \mathbb{Z}_p$, 
        adds $\{(g^{\alpha_0},B)\}$ to $L_H$, sets $(\hat{S},\hat{f}') = (\hat{g}^{s},\hat{R})$, with $s \leftarrow_\$ \mathbb{Z}_p$ and $\hat{R} \leftarrow_\$ \mathbb{G}_2$, and $(f,w, \tau') = (g^{\alpha_0},T, (\hat{S},\hat{f}'))$ and returns $(f,w, \tau',\pi_J,\sigma_{DS})$, where $\pi_J$ is the $\simulator$-simulated proof of knowledge $PK_J$ for $\alpha_0$ and $s$, and $\sigma_{DS}$ is the $DS$ signature on $\rho=e(g^{\alpha_0},\hat{g})$. Before returning, 
        $\mathcal{C}$ stores $\mathbf{msk}[i_0]=\alpha_0$.

        Finally, if $i=i_1$, $\challenger$ samples $(\alpha_1,\delta_1) \leftarrow_\$ \mathbb{Z}_p^2$, adds $\{(g^{\alpha_1},g^{\delta_1})\}$ to $L_H$, 
        sets $(\hat{S},\hat{f}') = (\hat{g}^{s},\hat{R})$, with $s \leftarrow_\$ \mathbb{Z}_p$ and $\hat{R} \leftarrow_\$ \mathbb{G}_2$, and $(f,w, \tau') = (g^{\alpha_1},A^{\delta_1}, (\hat{S},\hat{f}'))$ and returns $(f,w, \tau',\pi_J,\sigma_{DS})$, where $\pi_J$ is the $\simulator$-simu\-lated proof of knowledge $PK_J$ for $\alpha_1$ and $s$, and $\sigma_{DS}$ is the $DS$ signature on $\rho=e(g^{\alpha_1},\hat{g})$.
        Finally, $\mathcal{C}$ stores $\mathbf{msk}[i_1]=\alpha_1$.
        \item Trace: for an identity $i$ such that $i \neq i_0^* \land  i \neq i_1^*$, $\challenger$ executes the oracle algorithm, and aborts otherwise.
        \item Sig: $\challenger$ executes the oracle algorithm, except that the proof of knowledge of $\textbf{msk}[i]$ is simulated.

        \item Ch$_b$: let $(u,v,w)=\mathbf{mpk}[i_b]$ (if undefined, $\challenger$ returns $\perp$). $\challenger$ samples $r \leftarrow \mathbb{Z}_p$ and defines $nk = (u^r,v^r,w^r)$.
        He then $\simulator$-simulates the proof of knowledge $\sigma$ for $SPK_S$. He then updates $L_{ch} = L_{ch} \cup \{(i_0,m,nk,\sigma),(i_1,m,nk,\sigma)\}$  and returns $(nk,\sigma)$. 
    
    \end{itemize}

For user $i_0$, however, in the case where $T=g^{ab}$, the secret key $\alpha'_0$ for his $mpk$ (different from $\alpha_0$, as per experiment $G_6$) is the unknown $a=\log_gA$, since $u=H(f)=B$ and $w=B^{\alpha'_0}$, which must be equal to $T$. But, then, $a$ must also be the secret key $\alpha_1$ of honest user $i_1$, since, in that case, via $\SndToU$, for user $i_1$, we have $u=H(f)=g^{\delta_1}$ and $w=(g^{\delta_1})^{\alpha_1}= 
(g^{\delta_1})^{a}= A^{\delta_1}$. In this case, $\challenger$ simulates $G_F$; otherwise, the target users have different secret keys, and $\challenger$ simulates $G_6$.

\subsection{Compatibility of construction w.r.t. the NGS interface}
\label{sec:compatibility}
    That the construction of Section~\ref{sec:construction} indeed corresponds to the NGS
interface (see Section~\ref{sec:interface}) can be  seen by case analysis. As an example directly related to nicknames, we consider  the $\textit{Trace}$ case, i.e., the equivalence between $(nk \in [mpk]_\mathcal{R})$, for some $nk = (u', v', w')$ and issuer-generated $mpk = (u,v,w)$, and its construction $\textit{Trace}(ipk, \tau, nk)$, i.e., $(e(u',\tau)= e(w',\hat{g})) \land \GVf(ipk,nk)$, for $\tau$ compatible with $mpk$. 
    The forward direction is easy, since there exists $r$ such that $nk = mpk^r$; checking that  $\textit{Trace}(ipk, \tau, mpk^r)$ is true is straightforward, given $e$'s properties, and how $(u, v, w)$ is constrained in $Iss$, and $\tau$, in $Join$. 

    The backward direction assumes the existence of $nk = (u', v', w')$ such that, when passed to $\textit{Trace}$ with some $\tau$, the result is true. Since $\tau$ is in $\mathbb{G}_2$, there exists some $\alpha$, such that $\tau = \hat{g}^\alpha$. Since $e$ is injective, $w' = u'^\alpha$. Then, one can check that $v'$ satisfies the $Iss$-like constraint $v' = u'^x \cdot w'^y$. Completing the proof that $(nk \in [mpk]_\mathcal{R})$ thus just requires  finding some $r$ such that $u' = u^r$ (which would  also work for $v'$ and $w'$, which only depend on $u'$). Since both $u$ and $u'$ are in the cyclic group $\mathbb{G}_1$, of order $p$, there exists $i'$ and $i$ in $\mathbb{Z}_p$ such that $u' = g^{i'}$ and $u = g^i$. It suffices thus to pick 
    $r$ such that $i' \equiv ri ~ \textrm{mod}~p$, the existence of which is a direct consequence of B\'ezout's theorem, since $p$ is prime.

\end{document}